\documentclass[twocolumn,aps,showpacs,prl]{revtex4}
\usepackage{graphicx} 
\usepackage{color}
\usepackage{subfigure}
\usepackage{ulem}

\begin{document}
\title{Point defect concentrations in metastable Fe-C alloys}
\author{Clemens J. F\"orst,$^{1,2}$ 
        Jan Slycke,$^{3}$ 
        Krystyn J. Van Vliet,$^{2,*}$
        and Sidney Yip$^{1,2}$}
\affiliation{$^1$ Department of Nuclear Science and Engineering, Massachusetts Institute of Technology, Cambridge,
Massachusetts 02139, USA}
\affiliation{$^2$ Department of Materials Science and Engineering, Massachusetts Institute of Technology, Cambridge,
Massachusetts 02139, USA}
\affiliation{$^3$ Department of Materials Science, SKF Engineering Research Centre, P.O. Box 2350, 3430
DT Nieuwegein, The Netherlands}

\date{\today}
\begin{abstract}

Point defect species and concentrations in metastable Fe-C alloys
are determined using density functional theory and a constrained
free-energy functional. Carbon interstitials dominate unless iron
vacancies are in significant excess, whereas excess carbon causes greatly enhances
vacancy concentration.  Our predictions are amenable to experimental
verification; they provide a baseline for rationalizing complex
microstructures known in hardened and tempered steels, and by extension other
technological materials created by or subjected to extreme environments.

\end{abstract}


\maketitle

Many industrially significant alloys are intentionally processed as
metastable microstructures comprising a supersaturation of crystal
defects in various forms of aggregation ~\cite{Vehanen82, Ogawa02}. How
the defects are distributed and in what local concentrations they exist
are fundamental questions which are not yet resolved experimentally, yet
it is this microstructural complexity that governs the performance of
the material.  Hardened steels are an important example where
deformation behavior is intrinsically coupled to the lattice defects: in
this case, a body-centered cubic (bcc) Fe matrix with carbon content in
excess of equilibrium values for ferrite (solid solution of carbon in
bcc Fe) and cementite (Fe$_3$C), as well as high dislocation density and
supersaturation of vacancies~\cite{Vehanen82}. While it is generally
known that carbon binds strongly to crystal defects such as vacancies,
what is lacking is a robust computational strategy capable of linking
the fundamental physics of crystal defects and their energetics with
quantitative, microscopic details of the defect microstructure as it
evolves during processing and service.

In this work we determine the concentration of point defects and defect
clusters in Fe-C alloys using an approach based on first-principles
calculation of the formation energies of specific defects, and a
free-energy formulation allowing either the carbon or the iron vacancy
concentration to be out of equilibrium.  We find that the vacancy
concentration in the form of carbon-vacancy clusters increases
dramatically in the presence of carbon due to strongly exothermic
clustering reactions. Nevertheless, the concentrations of carbon-vacancy
clusters remain orders of magnitude below the concentration of
interstitial carbon. Only under the assumption of a significant
supersaturation of vacancies will carbon-vacancy clusters begin to
dominate the defect spectrum, in accord with experimental observation of
dominant carbon-vacancy clusters in irradiated bcc-Fe
alloys~\cite{Lavrentev98}.  Ab-initio results on the structure and
energetics of vacancies, self-interstitials and carbon interstitials in
bcc-Fe have been reported previously~\cite{Jiang03, Domain01}, as well
as the interaction of one and two C interstitials with the Fe
vacancies~\cite{Domain04}.  Our results at 0\,K are in close agreement
with cluster formation energies reported previously~\cite{Domain04}.
Additionally, we have considered clusters with higher carbon contents
and included di-vacancies in our considerations.

The supersaturation of carbon in such alloys has been established
through several experimental approaches, although the degree of
supersaturation in the bcc-Fe matrix ranges from 0.3\,at.\%\ C upon
400$^\circ$C annealing~\cite{Ohsaki05} to up to 5\,at.\% C in volumes
including ferritic grain boundaries~\cite{Wilde00, SKF}. While
experiments can quantify local carbon concentrations (e.g.,
three-dimensional atom probe~\cite{Wilde00}) or determine the
carbon-vacancy binding energy~\cite{McKee72} (a value of 0.85\,eV was
reported by positron annihilation measurements~\cite{Vehanen82}), it is
not yet feasible to measure the relative concentrations and specific
structures of C-vacancy clusters in bcc-Fe during thermal processing.
The most likely C-related defects are:  (1) simple interstitial
supersaturation~\cite{Ohsaki05}; (2) co-association with vacancies as
inferred from reduced vacancy diffusion~\cite{Vehanen82} and carbon
segregation to vacancy rich regions~\cite{Lavrentev98}; (3) direct
observation of elevated carbon concentration up to 8\,at.\% near
dislocation cores~\cite{Wilde00}; (4) up to 6.6\,at.\% at grain
boundaries~\cite{Wilde00, Ohsaki05}.   

We have performed a set of total energy calculations on the interaction
of carbon with single and double vacancies and applied the results in a
statistical mechanics model to obtain parameter-free estimates of the
concentration of point defects in various structural configurations in
the bcc-Fe matrix.  We employed density functional theory (DFT)
~\cite{Kohn,KohnSham,PBE} using Bl\"ochl's projector augmented wave
method~\cite{Blo94,Blo03} as implemented in the \textsc{VASP}
code~\cite{vasp2,vasppaw} with a plane-wave cutoff of 400\,eV.  The
calculations were performed in 128-atom supercells with the theoretical
lattice constant of 2.83\,\AA,  using a k-mesh of $2\times 2\times 2$
and a Methfessel Paxton Fermi-surface smearing parameter of
0.05\,eV~\cite{Methfessel89}.  No symmetry constraints were imposed.
Screening calculations and vibrational frequency calculations were
carried out in 54-atom supercells.  The geometry optimization was
terminated with a force cutoff of 5\,meV/\AA.  All calculations included
spin polarization, starting with a ferromagnetic charge density.
Because a wide variety of different bulk and defect structures are
considered, we anticipate an error of 0.1--0.2\,eV for the defect
formation energies.  At the target temperature of 160$^\circ$C, where
the body centered tetragonal phase transforms to bcc, this implies error in species
concentration of one to two orders of magnitude.  However, the overall
conclusions regarding the dominant defect concentrations are not found
to be affected by this level of uncertainty.

Table~\ref{tab:eform} summarizes the formation energies of the
defects and defect clusters under consideration,

\begin{eqnarray} \nonumber E^\mathrm{form}(T,\mu_{Fe}, \mu_C) &=&
E^\mathrm{D}(T) - E^\mathrm{0}(T)  \\ && - \mu_{Fe}\cdot \Delta n_{Fe} -
\mu_C\cdot \Delta n_C, \label{eqn:eform} \end{eqnarray} where
$E^\mathrm{D}(T)$ and $E^\mathrm{0}(T)$ denote the DFT energies of the
supercells with and without defects present, and $\mu_X$ and $\Delta
n_X$ are the chemical potential of species $X$ and the difference in
atom number between the two supercells.  Temperature dependence is taken
into account through configurational and vibrational free energy
contributions, the latter being calculated using the dynamical matrix of
all atoms affected by the defect formation. To approximate the
vibrational free energy we consider each atom as a three-dimensional
harmonic oscillator with force-constants derived from ab-initio total
energy calculations. The vibrational frequencies of carbon are found to
be strongly dependent on the defect geometry, varying between 7 and
34\,THz. Variation of the frequencies of Fe atoms in the vicinity of a
(double-) vacancy with respect to the bulk value is less than 3\,THz and
is therefore neglected.  The vibrational contributions to the formation
energies can amount to 0.2\,eV at 160$^\circ$C.

\begin{table}
\caption{Formation energies for the different defect stoichiometries at
160$^\circ$C. The chemical potentials $\mu_{Fe/C}$ (compare
Eqn.~\ref{eqn:eform}) are chosen to represent thermal equilibrium
of bulk Fe and Fe$_3$C.  $X_I$ and $X_V$ refer to an interstitial
and a vacancy of species $X$, respectively. Parentheses (\,\ldots)  denote a
defect cluster. The crystallographic directions [\,\ldots] refer to the
orientation of the iron double-vacancies.}
\label{tab:eform}
\vspace*{1mm}
\begin{tabular}{lr|lr}
\hline \hline
defect species & $E^\mathrm{form}$ [eV] & defect species & $E^\mathrm{form}$ [eV] \\ \hline
$C_I$ octahedral          & 0.58   &    $(2Fe_V + 2C)$ [100]      & 3.31   \\
$Fe_V$                    & 2.02   &    $(2Fe_V + 3C)$ [100]      & 3.21   \\
$Fe_I$                    & 3.91   &    $(2Fe_V + 4C)$ [100]      & 2.77   \\
$(Fe_V + 1C)$             & 1.96   &    $(2Fe_V + 5C)$ [100]      & 3.23   \\
$(Fe_V + 2C)$             & 1.53   &    $(2Fe_V + 6C)$ [100]      & 3.74   \\
$(Fe_V + 3C)$             & 1.98   &    $(2Fe_V + 7C)$ [100]      & 5.78   \\
$(Fe_V + 4C)$             & 3.03   &    $(2Fe_V + 1C)$ [111]      & 3.66   \\
$(Fe_V + 5C)$             & 6.58   &    $(2Fe_V + 2C)$ [111]      & 3.06   \\
$(Fe_V + 6C)$             & 13.48  &    $(2Fe_V + 3C)$ [111]      & 2.91   \\
$(2Fe_V)$ [100]           & 3.83   &    $(2Fe_V + 4C)$ [111]      & 2.43   \\
$(2Fe_V)$ [111]           & 3.85   &    $(2Fe_V + 5C)$ [111]      & 3.50   \\
$(2Fe_V + 1C)$ [100]      & 3.49   &    $(2Fe_V + 6C)$ [111]      & 4.76   \\
\hline \hline
\end{tabular}
\end{table}

Table~\ref{tab:eform} shows the most stable configuration for a given
stoichiometry. While there are different ways to arrange a given number
of C atoms around an Fe \mbox{(double-)}vacancy, we find that these
typically have very different formation energies.  A more extensive list
of energetics of one and two carbon atoms near a single vacancy can be
found in Ref.~\cite{Domain04}. Here we regard structures with formation
energies within 0.1\,eV (our approximate error) to be degenerate.
Structures with energies greater than than 0.3\,eV relative to the
lowest-energy configuration are not considered.

\begin{figure}
\centering
\includegraphics[width=4cm]{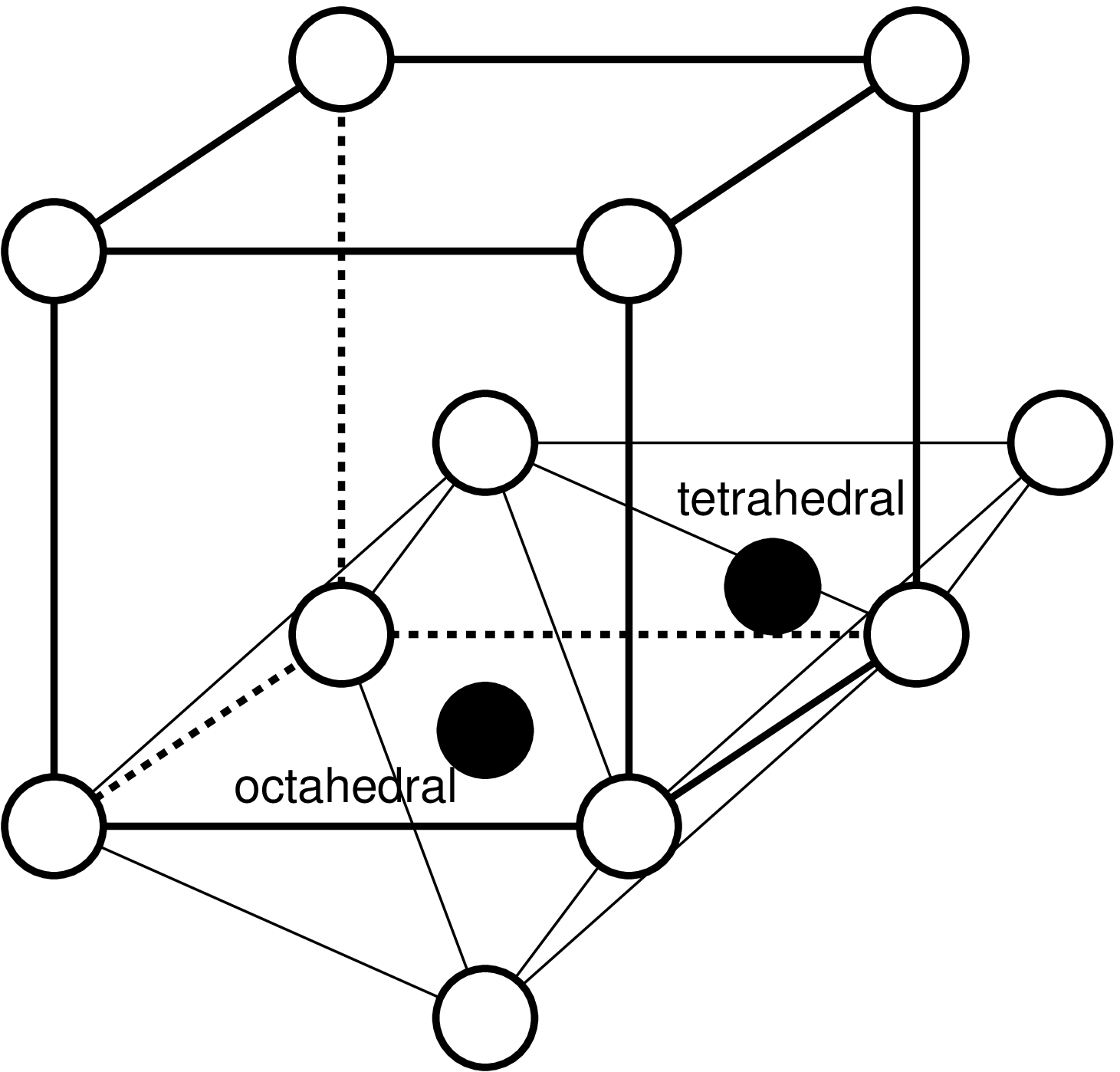}
\includegraphics[width=3.5cm]{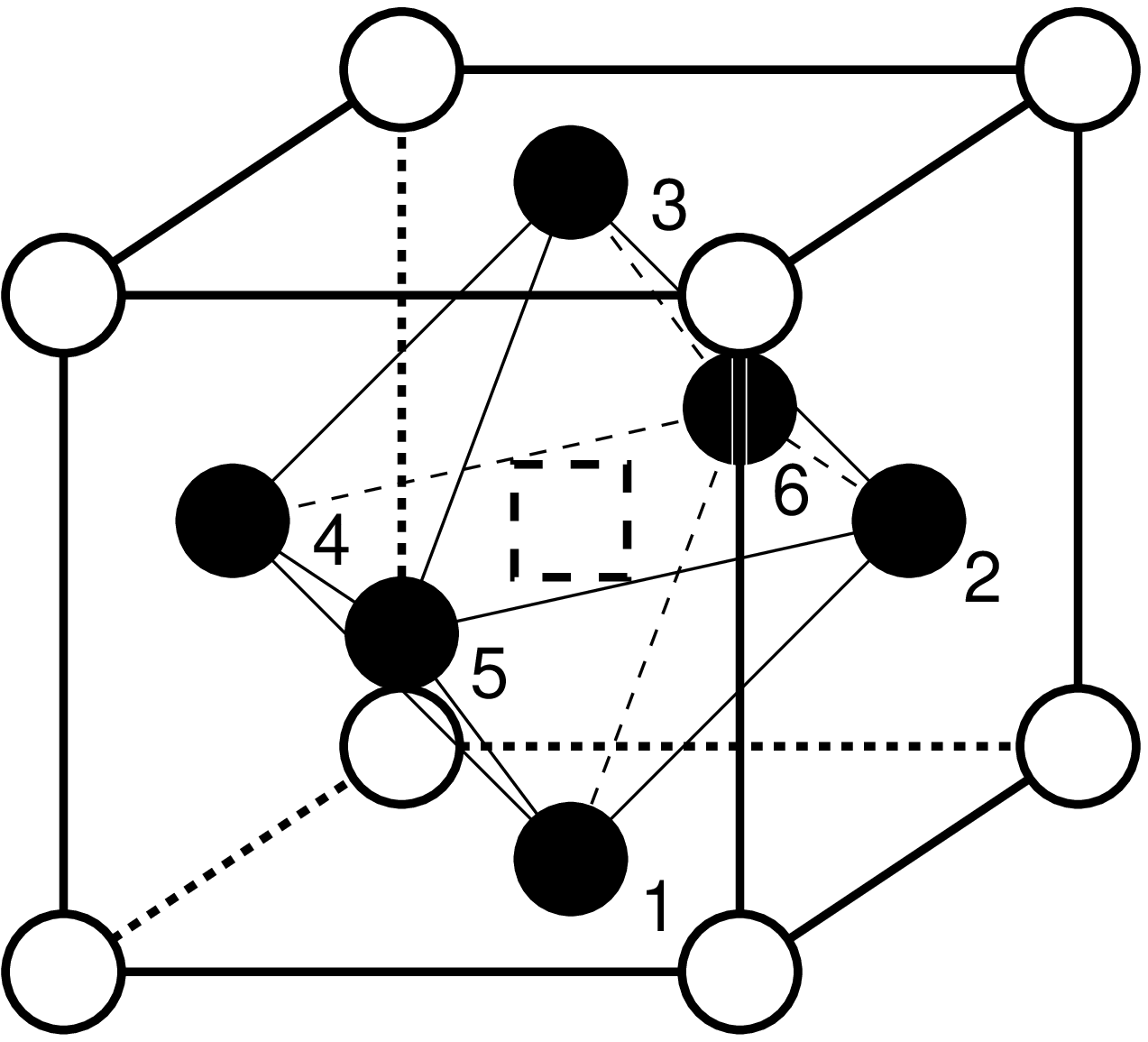} \\
\caption{Atomic structure of C interstitials and C-vacancy clusters in
bcc-Fe. Fe atoms in open circles,
carbon atoms filled circles. Left panel: octahedral and
tetrahedral carbon interstitials; right panel: position of carbon atoms
around a vacancy. A carbon-vacancy cluster with $n$ carbon atoms
contains the C atoms with labels from 1 to $n$.} 
\label{fig:defectstrc1}
\end{figure}

\begin{figure}
\centering
\includegraphics[height=5.2cm]{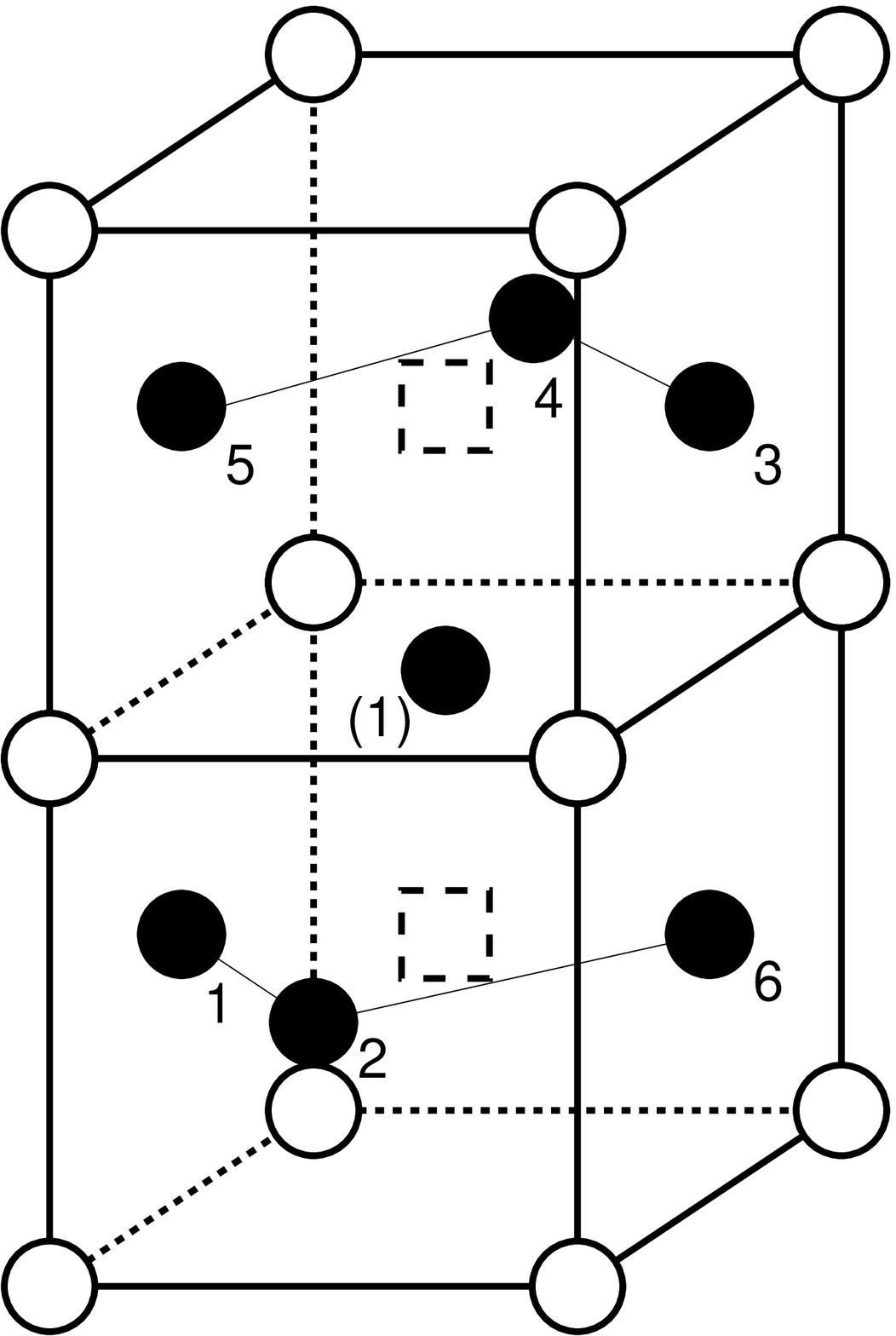}
\hfill
\includegraphics[height=4.5cm]{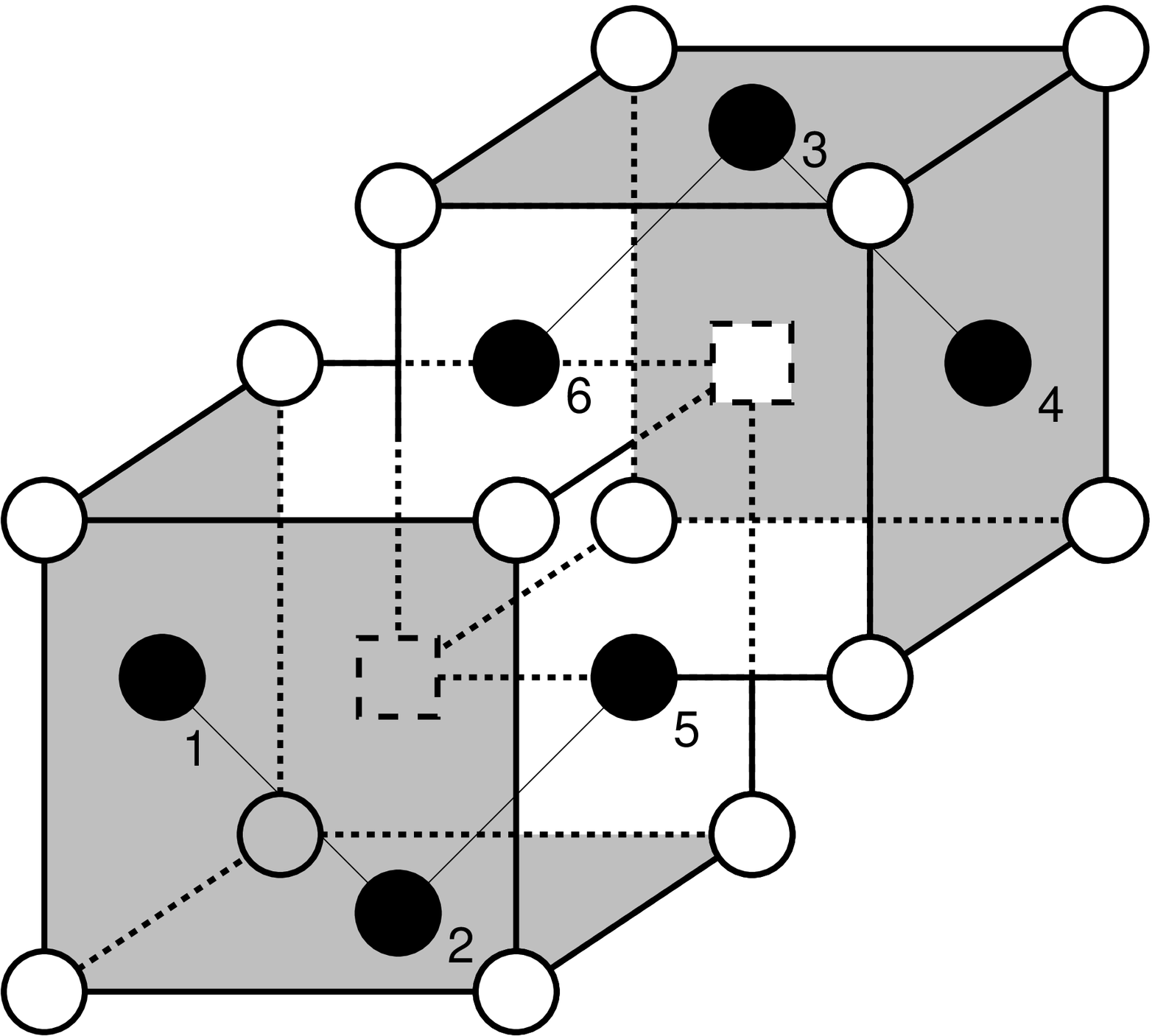}
\caption{Atomic structure of C-double vacancy clusters
oriented in [100] (left panel) and [111] (right panel) directions. Refer
to Fig.~\ref{fig:defectstrc1} for labeling. In case of the [100] vacancy,
the position '(1)' is only occupied if there is a single C atom.}
\label{fig:defectstrc2}
\end{figure}

Figure~\ref{fig:defectstrc1} shows the stable geometries for an Fe
vacancy surrounded by up to six carbon atoms. In the case of
di-vacancies, we need to distinguish between two possible orientations
[100] and [111], which are energetically practically degenerate in the
carbon-free state (compare Fig.~\ref{fig:defectstrc2} and
Table~\ref{tab:eform}).  Around a [111] vacancy there are two types of
adsorption sites: those with one of the six coordinating Fe atoms
missing (shaded sides in Fig.~\ref{fig:defectstrc2}) and those with two
missing coordinating atoms.  We find that configurations with two
missing coordinating Fe atoms are energetically unfavorable.  Otherwise,
the stable carbon geometries are analogous to that of the isolated
vacancy.  In the case of the [100] oriented di-vacancy, the configuration
of a single carbon atom situated between the two vacancies is most
favorable; however, this preference does not hold for pairs and triplets
of C atoms.  From screening a wide variety of C geometries, including
chains and tetrahedra connecting two vacancies, no new, low-energy
carbon configurations are found.

The formation of a C-vacancy cluster by combining a vacancy with $n$ carbon
interstitials,

\begin{equation} Fe_V + n\cdot C_I \longrightarrow (Fe_V + n\cdot C),
\end{equation} is seen to give a negative reaction enthalpy $\Delta H$
for \mbox{$n \le 4$} (Table~\ref{tab:eform}). The same is also predicted for
two single vacancies/C-vacancy clusters forming a
di-vacancy/C-di-vacancy cluster (e.g., \(2(Fe_V + 2C) \rightarrow (2Fe_V
+ 4C)\ [111]\) with $\Delta H = -0.6$\,eV). However, one should note
that the mass-action law for a clustering reaction such as $X+X
\rightarrow X_2$
relates the concentration of the cluster $c_{X_2}$ to the square of the
concentration of the individual defects, $(c_X)^2$.  Thus the defect
complex is the dominating species only if $\exp(-\frac{\Delta H}{k_BT})$
is larger than $1/c_X$. For typical defects in solids, $c_X$ is often
very small ($\ll$ 1\,ppm) and $\Delta H$ is on the order of $-0.1$\,eV, so
this condition is not necessarily fulfilled. Nonetheless, enthalpy
provides a measure of ``thermal stability" of a complex once it is
formed, as in the case of vacancy agglomerates~\cite{Vehanen82}.

The equilibrium concentrations of defects $c_i$ in a macroscopic crystal
are given by minimizing the free energy $F(T,\mu_{Fe},\mu_C)$ at given
temperature $T$ and chemical potentials $\mu_{Fe}$ and $\mu_C$:

\begin{eqnarray}
\nonumber
F(T,\mu_{Fe},\mu_C) &=& 
\sum_{i=1}^{M}
c_i  E^\mathrm{form}_i(T,\mu_{Fe},\mu_C) 
\\
 + k_BT &&\hspace*{-6mm} \sum_{i=1}^M \left[\vphantom{\sum} c_i\ln c_i +
(1-c_i)\ln (1-c_i) \right]
,
\label{eqn:efree}
\end{eqnarray}
where $E^\mathrm{form}$ is defined in Eqn.~\ref{eqn:eform}, $M$
denotes the number of different defect species and $c_i$ is the
concentration of defect species $i$~\cite{footnote1}.  One obtains from
Eqn.~\ref{eqn:efree}

\begin{equation}
c_i(T,\mu_{Fe},\mu_C) = \frac{1}{\exp\left(
\frac{E^\mathrm{form}_i(T,\mu_{Fe},\mu_C)}{k_B T}\right) +1 }.
  \label{eqn:conc}
\end{equation}

The dependence on the two chemical potentials, which enters through the
formation energies, can be turned into an explicit dependence on the
total carbon and/or vacancy concentrations if we impose the respective
constraints: 

\begin{equation} 
\sum_{i=1}^M c_i\cdot \Delta n^i_{Va/C} =
c^\mathrm{tot}_{Va/C}. 
\label{eqn:constraint}
\end{equation} 
This is useful for performing calculations under physically motivated
hypotheses.

As discussed above, carbon and vacancy concentrations in realistic
microstructures of Fe-C alloys need to be treated as nonequilibrium.  To
make the calculations tractable we consider two alternative hypotheses
under the condition of partial thermodynamic equilibrium. That is, we
have thermal equilibrium for one species (e.g., Fe) and non-equilibrium
for the other (e.g., C), thus reducing the problem to a study of the
effects of the concentration of the non-equilibrium species.  In
hypothesis (1) the carbon concentration is not in equilibrium with the
Fe$_3$C/Fe system, whereas Fe is assumed to be in equilibrated with bulk
Fe.  This scenario is motivated by the experimental observation that
carbide precipitation is kinetically hindered and requires tempering at
elevated temperatures~\cite{Ohsaki05}.   It is also implied that the Fe
matrix remains substantially undistorted.  In hypothesis (2) the vacancy
concentration is assumed to be non-equilibrium, whereas carbon is
assumed to equilibrate with the bulk Fe/Fe$_3$C system. This reflects
the anticipated processing conditions for hardening of Fe-C alloys; in
the course of martensitic transformation, large plastic strain leads to
high intrinsic defect concentrations, with vacancies binding strongly to
carbon interstitials.  Due to the increased diffusion barriers associated
with carbon-vacancy clusters~\cite{Vehanen82} and the absence of a
comparable number of Fe interstitials (high formation energies), it is
likely that a significant vacancy concentration is effectively frozen
in.  These two hypotheses represent extreme scenarios; any intermediate
situation can also be studied using the present data.

\begin{figure}
\centering
\includegraphics[width=7.5cm,clip=true]{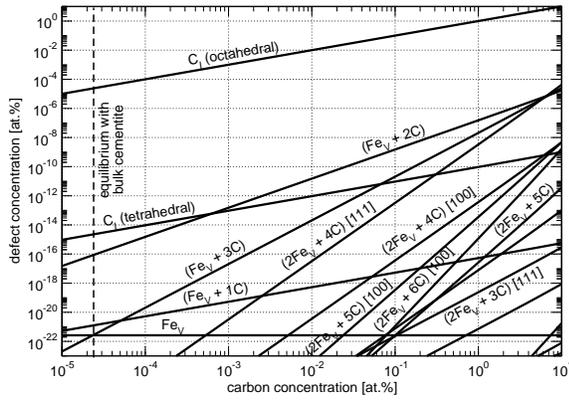}
\caption{Defect concentrations in a bcc-Fe matrix as a function of the total carbon
concentration at 160$^\circ$C. Labeling as in Table~\ref{tab:eform}. The vertical dashed
line marks thermal equilibrium with bulk cementite. Note that that small
precipitates are expected to have less favorable energetics in terms of atomic
and magnetic structure, which may shift the equilibrium carbon concentration to
higher values.}
\label{fig:ofcc}
\end{figure}

Figure~\ref{fig:ofcc} shows the consequences of hypothesis (1), the
variation of dominant defect species concentrations with the total
carbon concentration. These results are obtained by taking the chemical
potential $\mu_{Fe}$ to be the bulk Fe value, while $\mu_C$ is given by
the total carbon concentration (Eqn.~\ref{eqn:constraint}).  We observe
that the presence of carbon in the matrix can cause the vacancy
concentration (including carbon-vacancy clusters) to increase by 
15 orders of magnitude relative to the intrinsic vacancy
concentration ($Fe_V$) at a total carbon concentration of 1\,at.\%
Comparable increases in vacancy concentration have been reported in
the presence of a different impurity atom, hydrogen~\cite{Gavriljuk96,
McLellan97}. We also note that the concentration of carbon-free
di-vacancies is negligible. 

\begin{figure}
\centering
\includegraphics[width=7.5cm,clip=true]{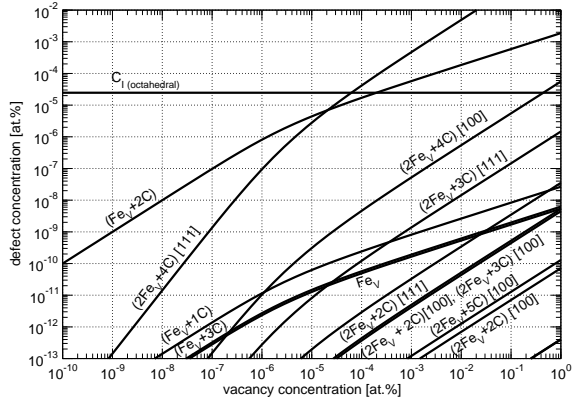}
\caption{Defect concentrations in a bcc-Fe matrix as a function of the total vacancy concentration 
at 160\,$^\circ$C. }
\label{fig:ofcv}
\end{figure}

Figure~\ref{fig:ofcv} shows the defect concentration variation with
total iron vacancy concentration under hypothesis (2).  In this case
$\mu_C$ is fixed at the reference value, the coexistence of Fe$_3$C and
bulk Fe.  We observe that the number of carbon atoms associated with
vacancies exceeds the number of carbon interstitials only when the
vacancy concentrations are larger than $10^{-4}$\,at.\%, which
corresponds to a typical equilibrium vacancy concentration near the
melting point~\cite{RHA94}. This correlates with the experimental
finding that C-vacancy clusters have been observed to be the dominant
carbon-species in irradiated samples~\cite{Lavrentev98}, which cannot be
explained under the assumption of an equilibrated vacancy concentration
(hypothesis (1)).

In conclusion, our work provides the first quantitative assessment of
the interplay of different defect species in Fe-C alloys. We predict
defect phase diagrams as a function of total carbon and vacancy
concentrations, results which are amenable to experimental verification.
As a computational framework for addressing microstructure complexity,
our approach should be applicable to other advanced technological
materials subject to extreme environmental conditions.

The authors thank Alejandro Sanz and Fred Lucas for stimulating
discussions, and acknowledge SKF for financial support. We have
benefited from computational resources funded by NSF (IMR-0414849).


%

\begin{thebibliography}{99}
\bibitem{footnote1} Note that we have assumed here that different defect
species do not compete for sites in the bcc lattice, i.e. the number of
iron sites available is large compared to the number of defects in the
system.  Considering a competition for defect sites the numerator of
Eqn.~\ref{eqn:conc} would be \((1 - \sum_{j\ne i} c_j)\).
\end{thebibliography}
\end{document}